\documentclass[aps,prc,showpacs,showkeys,nofootinbib]{revtex4}

\usepackage{graphicx}
\usepackage{dcolumn}
\usepackage{amsthm}
\usepackage{bm}

\newcommand{\definition}{\textit}

\newcommand{\rhobf}{\mbox{\boldmath $\rho$}}
\newcommand{\rhobfsm}{\small \mbox{\boldmath $\rho$}}

\begin{document}

\title
{Can a many-nucleon structure be visible in bremsstrahlung emission during $\alpha$ decay?}

\author{Sergei~P.~Maydanyuk}%
\email{maidan@kinr.kiev.ua}%
\affiliation{Institute of Modern Physics, Chinese Academy of Sciences, Lanzhou, 730000, China}
\affiliation{Institute for Nuclear Research, National Academy of Sciences of Ukraine, Kiev, 03680, Ukraine}
\author{Peng-Ming~Zhang}%
\email{zhpm@impcas.ac.cn} %
\affiliation{Institute of Modern Physics, Chinese Academy of Sciences, Lanzhou, 730000, China}
\author{Li-Ping~Zou}%
\email{zoulp@impcas.ac.cn} %
\affiliation{Institute of Modern Physics, Chinese Academy of Sciences, Lanzhou, 730000, China}











\date{\small\today}
\begin{abstract}
We analyze if the nucleon structure of the $\alpha$ decaying nucleus can be visible in the experimental bremsstrahlung spectra of the emitted photons which accompany such a decay.
We develop a new formalism of the bremsstrahlung model taking into account distribution of nucleons in the $\alpha$ decaying nuclear system.
We conclude the following:
(1) After inclusion of the nucleon structure into the model the calculated bremsstrahlung spectrum is changed very slowly for a majority of the $\alpha$ decaying nuclei. However, we have observed that visible changes really exist for the $^{106}{\rm Te}$ nucleus ($Q_{\alpha}=4.29$~MeV, $T_{1/2}$=70~mks) even for the energy of the emitted photons up to 1~MeV.
This nucleus is a good candidate for future experimental study of this task.
(2) Inclusion of the nucleon structure into the model increases the bremsstrahlung probability of the emitted photons.
(3) We find the following tendencies for obtaining the nuclei, which have bremsstrahlung spectra more sensitive to the nucleon structure: (a) direction to nuclei with smaller $Z$, (b) direction to nuclei with larger $Q_{\alpha}$-values.
\end{abstract}

\pacs{%
23.60.+e, 
41.60.-m, 
03.65.Xp, 
23.20.Js} 


\keywords{
bremsstrahlung,
photon,
alpha decay,
nucleon structure,
microscopic model,
Dirac equation,
Pauli equation,
tunneling}

\maketitle

\section{Introduction
\label{sec.inroduction}}

The bremsstrahlung emission of photons accompanying nuclear reactions has been causing much interest for a long time
(see reviews~\cite{Pluiko.1987.PEPAN,Kamanin.1989.PEPAN} and
books~\cite{Amusia_Buimistrov.1987,Amusia.1990,Maydanyuk.2012.book_bremsstrahlung}).
This is because of such photons provide rich information about the studied nuclear process.
Dynamics of the nuclear process, interactions between nucleons, types of nuclear forces, quantum effects, anisotropy (deformations) can be included in the model describing the bremsstrahlung emission.
At the same time, measurements of such photons and their further analysis provide an evaluation of the suitability of the components of the model.

However, progress using dynamics and interactions between nucleons and nuclear forces has been limited.
Researchers pointed to such a difficulty, who included the realistic potentials of nuclear interactions in calculations of the bremsstrahlung spectra (for example, see \cite{Kopitin.1997.YF}).
This is also reflected by the small number of papers in this research area.
We explain such a situation by the essential distance between
(1) achievement of good agreement of the existing experimental data with the calculated spectra, and
(2) development of mathematical formalism, sufficiently sensitive to many-nucleon interactions and dynamics, which should give convergent calculations of the spectra and explain the experimental data.

Essential efforts were made to understand emission of the bremsstrahlung photons in nucleon-nucleon and nucleon-nuclear collisions.
But, a prevailing idea of the existed models consists in reduction of the complicated interactions between nuclei to two-nucleon interactions, which are assumed to be leading.
However, main emphasis in such papers was made on construction of correct relativistic description of interaction between two nucleons in this task, where formalism was developed mainly in momentum representation.
Here, we should like to note two directions of intensive investigations:
~\cite{Nakayama.1989.PRC,Herrmann.1991.PRC} and
~\cite{Liou.1987.PRC,Liou.1993.PRC,Liou.1995.PLB.v345,Liou.1995.PLB.v355,Liou.1996.PRC,Li.1998.PRC.v57,Li.1998.PRC.v58,%
Timmermans.2001.PRC,Liou.2004.PRC,Li.2005.PRC,Timmermans.2006.PRC,Li.2011.PRC}.


The bremsstrahlung of the emitted photons in nuclear reactions where the studied nuclei were described on the microscopic level, was previously studied.
Of course, the advances made by 
Baye, Descouvemont, Keller, Sauwens, Liu, Tang, Kanada, Dohet-Eraly and Sparenberg%
~\cite{Baye.1985.NPA,Baye.1991.NPA,Liu.1990.PRC,
Dohet-Eraly.2013.PhD,Dohet-Eraly.2011.PRC,Dohet-Eraly.2011.JPCS,Dohet-Eraly.2013.JPCS,Dohet-Eraly.2013.PRC}.
need to be noted.
%
The idea of a many-nucleon description of the nuclei in this task appears attractive.
It could be interesting to develop this for reactions with participation of heavy nuclei,
where there is more evidence
(for example, see
research papers~\cite{Edington.1966.NP,Koehler.1967.PRL,Kwato_Njock.1988.PLB,%
Pinston.1989.PLB,Pinston.1990.PLB,Clayton.1992.PRC},
reviews~\cite{Pluiko.1987.PEPAN,Kamanin.1989.PEPAN}
and PhD thesis~\cite{Clayton.1991.PhD}
for the bremsstrahlung during scattering of protons off nuclei for low and intermediate energies of the emitted photons
and theoretical description in~\cite{Maydanyuk.2012.PRC,Maydanyuk_Zhang.2015.PRC,Nakayama.1989.PRC,Nakayama.1986.PRC,Remington.1987.PRC},
~\cite{D'Arrigo.1994.PHLTA,Kasagi.1997.JPHGB,Kasagi.1997.PRLTA,%
Boie.2007.PRL,Maydanyuk.2008.EPJA,Giardina.2008.MPLA} for the bremsstrahlung during the $\alpha$ decay of nuclei
and fully quantum calculations in~\cite{Maydanyuk.2006.EPJA,Maydanyuk.2008.EPJA,Giardina.2008.MPLA,Papenbrock.1998.PRLTA,Tkalya.1999.PHRVA}
and semiclassical calculations in \cite{Jentschura.2008.PRC},
\cite{Maydanyuk.2009.NPA} for extraction of information about nuclear deformations of the $\alpha$  decaying nuclei via analysis of the experimental bremsstrahlung spectra,
\cite{Ploeg.1995.PRC,Kasagi.1989.JPSJ,Luke.1991.PRC,Varlachev.2007.BRASP,%
Hofman.1993.PRC,Eremin.2010.IJMPE,Pandit.2010.PLB} for the bremsstrahlung in fission of heavy nuclei and theoretical description in \cite{Maydanyuk.2010.PRC},
also \cite{Maydanyuk.2011.JPG} for predictions of the bremsstrahlung during emission of proton from nuclei).
%
%
A more empirical evidence provides us more possibilities for testing the developed models.
However, the first question which should be clarified is how much the nucleon structure of nuclei and incident fragments are visible in the bremsstrahlung spectra.
It may appear that inclusion of the many-nucleon structure into the model will barely change visibly the calculated spectra for energies and parameters used in experiments.
Moreover, the development of the microscopic formalism for such a problem is difficult.
So, before developing an accurate model, its practicality needs to be clarified.

This paper is devoted to analysis and solution of such a question, where the $\alpha$ decay is chosen as the studied reaction.
We explain such a choice by the following.
(1) The microscopic approaches can be applied for description of the $\alpha$ decaying nuclei by natural means (for example, see some research papers \cite{Thomas.1954.PTP,Delion.1992.PRC,Xu.2006.PRC,Silisteanu.2012.ADNDT,Delion.2013.PRC}, reviews \cite{Silisteanu.2012.ADNDT,Ivascu.1990.PEPAN,Lovas.1998.PRep,Hodgson.2003.PRep} and reference therein).
(2) The $\alpha$ decaying nuclei are heavy.
(3) Progress has been achieved in study of the $\alpha$ nucleus interactions tested experimentally, that makes $\alpha$ decay one of the most deeply studied reactions in nuclear physics.
(4) Rich material has been developed in theoretical description of the bremsstrahlung emission of photons in the given reaction~\cite{Batkin.1986.SJNCA,Dyakonov.1996.PRLTA,Kasagi.1997.JPHGB,Kasagi.1997.PRLTA,
Papenbrock.1998.PRLTA,Dyakonov.1999.PHRVA,Bertulani.1999.PHRVA,Takigawa.1999.PHRVA,Flambaum.1999.PRLTA,
Tkalya.1999.JETP,Tkalya.1999.PHRVA,So_Kim.2000.JKPS,Misicu.2001.JPHGB,Dijk.2003.FBSSE,
Maydanyuk.2003.PTP,Maydanyuk.2006.EPJA,Ohtsuki.2006.CzJP,
Amusia.2007.JETP,Jentschura.2008.PRC,Maydanyuk.2008.EPJA,
Giardina.2008.MPLA,Maydanyuk.2009.NPA,Maydanyuk.2009.TONPPJ,Maydanyuk.2009.JPS}.
(5) In study of the bremsstrahlung photons during the $\alpha$ decay the close agreement between theory and existing experimental data has been achieved~\cite{Kasagi.1997.JPHGB,Kasagi.1997.PRLTA,Maydanyuk.2008.EPJA,Giardina.2008.MPLA,D'Arrigo.1994.PHLTA,Boie.2007.PRL}.
(6) Such an agreement (for example, see~\cite{Maydanyuk.2009.TONPPJ} for details and demonstrations) is obtained without normalization of the calculated spectra on the experimental data (that has been very rarely achieved for other reactions with good accuracy), which is a significant advance for new predictions and inclusion of all sets of the $\alpha$ decaying nuclei into analysis ($\alpha$ decay is observed for more than 420 nuclei with $A > 105$ and $Z > 52$).


\section{Model
\label{sec.2.2}}

\subsection{Operator of emission of the $\alpha$ nucleus system
\label{sec.2.2}}

We shall start from the leading form (7) of the photon emission operator $\hat{H}_{\gamma}$ in \cite{Maydanyuk_Zhang.2015.PRC}, generalizing it for the system of an $\alpha$ particle composed from 4 nucleons and daughter nucleus composed of $A$ nucleons in the laboratory system.
Using presentation for the vector potential of the electromagnetic field in form (5) in \cite{Maydanyuk.2012.PRC},
we obtain
\begin{equation}
\begin{array}{lcl}
  \hat{H}_{\gamma} =
    -\,e\, \sqrt{\displaystyle\frac{2\pi}{w_{\rm ph}}}\,
    \displaystyle\sum\limits_{\alpha=1,2} \mathbf{e}^{(\alpha),*}\;
    \biggl\{
      \displaystyle\sum\limits_{i=1}^{4}
        \displaystyle\frac{z_{i}}{m_{i}}\; e^{-i \mathbf{kr}_{i}}\, \mathbf{p}_{i} +
      \displaystyle\sum\limits_{j=1}^{A}
        \displaystyle\frac{z_{j}}{m_{j}}\; e^{-i \mathbf{kr}_{j}}\, \mathbf{p}_{j}
    \Bigr\}.
\end{array}
\label{eq.2.2.3}
\end{equation}
%
%
%
%
%
%
Here,
star denotes complex conjugation,
$z_{s}$ and $m_{s}$ are the electromagnetic charge and mass of the nucleon with number $s$,
$\mathbf{p}_{s} = -i\hbar\, \mathbf{d}/\mathbf{dr}_{s} $ is the momentum operator for the nucleon with number $s$
($s=i$ is for nucleon of the $\alpha$ particle and $s=j$ for nucleon of the daughter nucleus;
we number nucleons of $\alpha$ particle by index $i$, and nucleons of the nucleus by index $j$).
$\mathbf{e}^{(\alpha)}$ are unit vectors of the polarization of the photon emitted [$\mathbf{e}^{(\alpha), *} = \mathbf{e}^{(\alpha)}$], $\mathbf{k}$ is the wave vector of the photon and $w_{\rm ph} = k c = \bigl| \mathbf{k} \bigr|\: c$. Vectors $\mathbf{e}^{(\alpha)}$ are perpendicular to $\mathbf{k}$ in the Coulomb gauge. We have two independent polarizations $\mathbf{e}^{(1)}$ and $\mathbf{e}^{(2)}$ for the photon with momentum $\mathbf{k}$ ($\alpha=1,2$).
In this paper we shall use the system of units where $\hbar = 1$ and $c = 1$.

Now we turn to the center-of-mass frame. We define coordinate of centers of masses for the $\alpha$ particle  as $\mathbf{r}_{\alpha}$, for the daughter nucleus as $\mathbf{R}_{A}$ and
for the complete system as $\mathbf{R}$ having form
$\mathbf{r}_{\alpha} = \sum_{i=1}^{4} m_{i}\, \mathbf{r}_{\alpha i} / m_{\alpha}$,
$\mathbf{R}_{A} = \sum_{j=1}^{A} m_{j}\, \mathbf{r}_{A j} / m_{A}$,
$\mathbf{R} = (m_{A}\mathbf{R}_{A} + m_{\alpha}\mathbf{r}_{\alpha}) / (m_{A}+m_{\alpha})$,
%
%
%
where $m_{\alpha}$ and $m_{A}$ are masses of the $\alpha$ particle and daughter nucleus.
Introducing new relative coordinates $\rhobf_{\alpha i}$, $\rhobf_{A j}$ and $\mathbf{r}$ as
$\mathbf{r}_{i} = \mathbf{r}_{\alpha} + \rhobf_{\alpha i}$,
$\mathbf{r}_{j} = \mathbf{R}_{A} + \rhobf_{A j}$,
$\mathbf{r} = \mathbf{r}_{\alpha} - \mathbf{R}_{A}$,
%
%
we find the corresponding momenta
$\mathbf{p}_{i} = \mathbf{p}_{\alpha} + \mathbf{\tilde{p}}_{\alpha i}$,
$\mathbf{p}_{j} = \mathbf{P}_{A} + \mathbf{\tilde{p}}_{A j}$,
$\mathbf{p} = \mathbf{p}_{\alpha} - \mathbf{P}_{A}$,
%
%
where
$\mathbf{p}_{\alpha} = -\,i\hbar\, \mathbf{d} / \mathbf{dr}_{\alpha}$,
$\mathbf{\tilde{p}}_{\alpha i} = -\,i\hbar\, \mathbf{d} / \mathbf{d\rho}_{\alpha i}$,
$\mathbf{P}_{A} = -\,i\hbar\, \mathbf{d} / \mathbf{dR}_{A}$,
$\mathbf{\tilde{p}}_{A j} = -i\hbar\, \mathbf{d} / \mathbf{d\rho}_{A j}$.
Using these formulas, we obtain
\begin{equation}
\begin{array}{cccc}
  \mathbf{R}_{A} = \mathbf{R} - c_{\alpha}\, \mathbf{r}, &
  \mathbf{r}_{\alpha} = \mathbf{R} + c_{A}\, \mathbf{r}, %
  \mathbf{r}_{i} = \mathbf{R} + c_{A}\, \mathbf{r} + \rhobf_{\alpha i}, &
  \mathbf{r}_{j} = \mathbf{R} - c_{\alpha}\, \mathbf{r} + \rhobf_{A j},
\end{array}
\label{eq.2.2.7}
\end{equation}
%
%
%
where we introduced $c_{A} = \frac{m_{A}}{m_{A}+m_{\alpha}}$ and $c_{\alpha} = \frac{m_{\alpha}}{m_{A}+m_{\alpha}}$.
Substituting these expressions to Eq.~(\ref{eq.2.2.3}), we find ($m_{\rm p}$ is mass of proton)%
\footnote{As we have only 3 independent variables $\rhobf_{\alpha i}$ and $A-1$ independent variables $\rhobf_{Aj}$,
Eq.~(\ref{eq.2.2.9}) can be rewritten without variables $\rhobf_{\alpha 4}$, $\rhobf_{AA}$ and $\mathbf{\tilde{p}}_{\alpha 4}$, $\mathbf{\tilde{p}}_{AA}$.}
\begin{equation}
\begin{array}{lcl}
  \vspace{0mm}
  \hat{H}_{\gamma} \;\; = \;\;
  -\,e\; \sqrt{\displaystyle\frac{2\pi}{w_{\rm ph}}}\,
    \displaystyle\sum\limits_{\alpha=1,2} \mathbf{e}^{(\alpha),*}\;
    e^{-i \mathbf{k} \bigl[\mathbf{R} - c_{\alpha} \mathbf{r} \bigr]}\; 
    \Biggl\{
      \Bigl[
        e^{-i \mathbf{k}\mathbf{r}}\,
          \displaystyle\sum\limits_{i=1}^{4}  \displaystyle\frac{z_{i}}{m_{i}}\;
          e^{-i \mathbf{k} \rhobf_{\alpha i} }\, +\,

          \displaystyle\sum\limits_{j=1}^{A}  \displaystyle\frac{z_{j}}{m_{j}}\;
          e^{-i \mathbf{k} \rhobfsm_{A j} }
      \Bigr]\: \mathbf{P}\; + \\

  \;\; +\:
      \Bigl[
        c_{A}\, e^{-i \mathbf{k}\mathbf{r}}
          \displaystyle\sum\limits_{i=1}^{4} \displaystyle\frac{z_{i}}{m_{i}}\;
          e^{-i \mathbf{k} \rhobf_{\alpha i} }\, -
        c_{\alpha}
          \displaystyle\sum\limits_{j=1}^{A} \displaystyle\frac{z_{j}}{m_{j}}\;
          e^{-i \mathbf{k} \rhobf_{A j} }
      \Bigr]\: \mathbf{p}\; + 

      e^{-i \mathbf{k}\mathbf{r}}
      \displaystyle\sum\limits_{i=1}^{4}
        \displaystyle\frac{z_{i}}{m_{i}}\;
        e^{-i \mathbf{k} \rhobf_{\alpha i} }\, \mathbf{\tilde{p}}_{\alpha i}\: +\:
      \displaystyle\sum\limits_{j=1}^{A}
        \displaystyle\frac{z_{j}}{m_{j}}\;
        e^{-i \mathbf{k} \rhobf_{A j}}\, \mathbf{\tilde{p}}_{A j}
    \Biggr\}.
\end{array}
\label{eq.2.2.9}
\end{equation}

\subsection{Wave function of the $\alpha$-nucleus system
\label{sec.2.3}}

Emission of the bremsstrahlung photons is caused by the relative motion of nucleons of the full nuclear system. However, as the most intensive emission of photons is formed by relative motion of the $\alpha$ particle related to the nucleus, it is sensible to represent the total wave function via coordinates of relative motion of these complicated objects.
In this paper we follow the formalism given in \cite{Maydanyuk_Zhang.2015.PRC} for the proton-nucleus scattering, and we add description of many-nucleon structure of the $\alpha$-particle.
Such a presentation of the wave function allows us to take into account the most accurately the leading contribution of the wave function of relative motion into the bremsstrahlung spectrum, while the many nucleon structure of the $\alpha$ particle and nucleus should provide only minor corrections (such a contribution of the many-nucleon structure follows from good agreement between theory and experiment for $\alpha$ decay obtained without the many-nucleon structure,
see \cite{Maydanyuk.2006.EPJA,Maydanyuk.2008.EPJA,Giardina.2008.MPLA,Maydanyuk.2009.NPA}).
Before developing a detailed many-nucleon formalism for such a problem, we shall clarify first if the many-nucleon structure of the $\alpha$ nucleus system is visible in the experimental bremsstrahlung spectra.
In this regard, estimation of many-nucleon contribution in the full bremsstrahlung spectrum is well described task.
Thus, we define the wave function of the full nuclear system as
\begin{equation}
  \Psi =
  \Phi (\mathbf{R}) \cdot
  \Phi_{\rm \alpha - nucl} (\mathbf{r}) \cdot
  \psi_{\rm nucl} (\beta_{A}) \cdot
  \psi_{\alpha} (\beta_{\alpha}),
\label{eq.2.3.1}
\end{equation}
where
\begin{equation}
\begin{array}{lcl}
  \psi_{\rm nucl} (\beta_{A}) =
  \psi_{\rm nucl} (1 \cdots A ) =
  \displaystyle\frac{1}{\sqrt{A!}}
  \displaystyle\sum\limits_{p_{A}}
    (-1)^{\varepsilon_{p_{A}}}
    \psi_{\lambda_{1}}(1)
    \psi_{\lambda_{2}}(2) \ldots
    \psi_{\lambda_{A}}(A), \\

  \psi_{\alpha} (\beta_{\alpha}) =
  \psi_{\alpha} (1 \cdots 4) =
  \displaystyle\frac{1}{\sqrt{4!}}
  \displaystyle\sum\limits_{p_{\alpha}}
    (-1)^{\varepsilon_{p_{\alpha}}}
    \psi_{\lambda_{1}}(1)
    \psi_{\lambda_{2}}(2)
    \psi_{\lambda_{3}}(3)
    \psi_{\lambda_{4}}(4).
\end{array}
\label{eq.2.3.3}
\end{equation}
Here, $\beta_{\alpha}$ is the set of numbers $1 \cdots 4$ of nucleons of the $\alpha$ particle,
$\beta_{A}$ is the set of numbers $1 \cdots A$ of nucleons of the nucleus,
$\Phi (\mathbf{R})$ is the function describing motion of center-of-mass of the full nuclear system,
$\Phi_{\rm \alpha - nucl} (\mathbf{r})$ is the function describing relative motion of the $\alpha$ particle concerning to nucleus (without description of internal relative motions of nucleons in the $\alpha$ particle and nucleus),
$\psi_{\alpha} (\beta_{\alpha})$ is the many-nucleon function dependent on nucleons of the $\alpha$ particle (it determines space state on the basis of relative distances $\rhobf_{1} \cdots \rhobf_{4}$ of nucleons of the $\alpha$ particle concerning to its center-of-mass),
$\psi_{\rm nucl} (\beta_{A})$ is the many-nucleon function dependent on nucleons of the nucleus.
Summation in Eqs.~(\ref{eq.2.3.3}) is performed over all $A!$ permutations of coordinates or states of nucleons. One-nucleon functions $\psi_{\lambda_{s}}(s)$ represent the multiplication of space and spin-isospin
functions as
$\psi_{\lambda_{s}} (s) = \varphi_{n_{s}} (\mathbf{r}_{s})\, \bigl|\, \sigma^{(s)} \tau^{(s)} \bigr\rangle$,
%
%
where
$\varphi_{n_{s}}$ is the space function of the nucleon with number $s$,
$n_{s}$ is the number of state of the space function of the nucleon with number $s$,
$\bigl|\, \sigma^{(s)} \tau^{(s)} \bigr\rangle$ is the spin-isospin function of the nucleon with number $s$.
%

We include the many-nucleon structure into wave functions $\psi_{\rm nucl}$ and $\psi_{\alpha}$ of nucleus and $\alpha$ particle while we assume that wave function of relative motion $\Phi_{\rm \alpha - nucl} (\mathbf{r})$ is calculated without them but with maximal orientation of the $\alpha$ nucleus potential extracted from experimental data of $\alpha$ decay, $\alpha$ nucleus scattering and $\alpha$ capture.
So, $\psi_{\alpha}$ and $\psi_{\rm nucl}$ describe only the internal states of the $\alpha$ particle and nucleus. Motion of nucleons of the nucleus relative to each other does not influence on the internal states of the $\alpha$ particle and, therefore, such a representation of the wave function can be considered as an approximation. However, both relative internal motions of nucleons of the $\alpha$ particle and the nucleus provide their contributions to the full bremsstrahlung spectrum and can be estimated. In such a sense we take into account the internal nucleon structure of the $\alpha$ particle and
nucleus.
We calculate the matrix element of the photon emission as
\begin{equation}
\begin{array}{lcl}
\vspace{1mm}
  \langle \psi_{f} (1 \cdots A )\, |\, \hat{H}_{\gamma}\, |\, \psi_{i} (1 \cdots A ) \rangle =  \\
 \vspace{1mm}
   = \quad
    \displaystyle\frac{1}{A\,(A-1)}\;
    \displaystyle\sum\limits_{k=1}^{A}
    \displaystyle\sum\limits_{m=1, m \ne k}^{A}
    \biggl\{
      \langle \psi_{k}(i)\, \psi_{m}(j) |\, \hat{H}_{\gamma}\, |\, \psi_{k}(i)\, \psi_{m}(j) \rangle - 

    \langle \psi_{k}(i)\, \psi_{m}(j) |\, \hat{H}_{\gamma}\, |\, \psi_{m}(i)\, \psi_{k}(j) \rangle
  \biggr\}.
\end{array}
\label{eq.2.3.5}
\end{equation}

\subsection{Matrix element of emission and effective charge
\label{sec.2.4}}

We shall assume
$\Phi_{\bar{s}} (\mathbf{R}) =  e^{-i\,\mathbf{K}_{\bar{s}}\cdot\mathbf{R}}$
%
%
%
where $\bar{s} = i$ or $f$ (indexes $i$ and $f$ denote the initial state, i.e. the state before emission of photon,
and the final state, i.e. the state after emission of photon),
$\mathbf{K}_{s}$ is momentum of the total system~\cite{Kopitin.1997.YF}.
Suggesting
%
$
  \mathbf{K}_{i} = 0,
$
%
we calculate the matrix element:
\begin{equation}
\begin{array}{lcl}
  \vspace{0mm}
  \langle \Psi_{f} |\, \hat{H}_{\gamma} |\, \Psi_{i} \rangle  \;\; = \;\;
  -\,\displaystyle\frac{e}{m_{\rm p}}\; \sqrt{\displaystyle\frac{2\pi}{w_{\rm ph}}}\,
    \displaystyle\sum\limits_{\alpha=1,2} \mathbf{e}^{(\alpha),*}\;
  \Bigl\{ M_{1} + M_{2} + M_{3} + M_{4} \Bigr\},
\end{array}
\label{eq.2.4.2}
\end{equation}
where
\begin{equation}
\begin{array}{lcl}
  M_{1} =
  \biggl\langle
    \Psi_{f}\,
  \biggl|\,
    e^{i\,(\mathbf{K}_{f} - \mathbf{k})\cdot\mathbf{R}}\:
    e^{i\, c_{\alpha} \mathbf{kr} }\; 

      \Bigl[
        e^{-i \mathbf{k}\mathbf{r}}\,
          \displaystyle\sum\limits_{i=1}^{4} z_{i}\, \displaystyle\frac{m_{\rm p}}{m_{i}}\;
          e^{-i \mathbf{k} \rhobf_{\alpha i} }\, +
          \displaystyle\sum\limits_{j=1}^{A} z_{j}\, \displaystyle\frac{m_{\rm p}}{m_{j}}\;
          e^{-i \mathbf{k} \rhobf_{A j} }
      \Bigr]\: \mathbf{P}\;
    \biggr|\,
      \Psi_{i}
    \biggr\rangle, \\
%
  M_{2} =
    \biggl\langle
      \Psi_{f}\,
    \biggl|\,
      e^{i\,(\mathbf{K}_{f} - \mathbf{k})\cdot\mathbf{R}}\:
      e^{i\, c_{\alpha} \mathbf{kr} }\,
      \Bigl[
          e^{-i \mathbf{k}\mathbf{r}}\, c_{A}\,
          \displaystyle\sum\limits_{i=1}^{4} z_{i}\,\displaystyle\frac{m_{\rm p}}{m_{i}}\;
          e^{-i \mathbf{k} \rhobf_{\alpha i} }\, - 

        c_{\alpha}\,
          \displaystyle\sum\limits_{j=1}^{A} z_{j}\,\displaystyle\frac{m_{\rm p}}{m_{j}}\;
          e^{-i \mathbf{k} \rhobf_{A j} }
      \Bigr]\: \mathbf{p}
    \biggr|\,
      \Psi_{i}\,
    \biggr\rangle, \\
%
  M_{3} =
    \biggl\langle
      \Psi_{f}
    \biggl|
      e^{i(\mathbf{K}_{f} - \mathbf{k})\cdot\mathbf{R}}\:
      e^{i\, c_{\alpha} \mathbf{kr} }
      e^{-i\, \mathbf{kr}}\,
      \displaystyle\sum\limits_{i=1}^{4}
        z_{i}\,\displaystyle\frac{m_{\rm p}}{m_{i}}\;
        e^{-i \mathbf{k} \rhobf_{\alpha i} }\, \mathbf{\tilde{p}}_{\alpha i}
    \biggr|
      \Psi_{i}
    \biggr\rangle, \\
%
  M_{4} =
    \biggl\langle
      \Psi_{f}\,
    \biggl|\,
      e^{i\,(\mathbf{K}_{f} - \mathbf{k})\cdot\mathbf{R}}\:
      e^{i\, c_{\alpha} \mathbf{kr} }\,
      \displaystyle\sum\limits_{j=1}^{A}
        z_{j}\,\displaystyle\frac{m_{\rm p}}{m_{j}}\;
        e^{-i \mathbf{k} \rhobf_{A j}}\, \mathbf{\tilde{p}}_{A j}\;
    \biggr|\,
      \Psi_{i}
    \biggr\rangle.
\end{array}
\label{eq.2.4.2}
\end{equation}
%
We will not use the first term $M_{1}$ (as we shall study decay in the center-of-mass system and neglect by possible response), the third term $M_{3}$ (as we shall neglect by the contribution of photon emission caused by the deformation of the $\alpha$ particle as it leaves) and the forth term $M_{4}$ (as we shall not study the contribution of photon emission caused by the deformation of the daughter nucleus during decay).
On this basis we obtain:
\begin{equation}
\begin{array}{lcl}
\vspace{1mm}
  \langle \Psi_{f} |\, \hat{H}_{\gamma} |\, \Psi_{i} \rangle & = &
  -\,\displaystyle\frac{e}{m_{\rm p}}\; \sqrt{\displaystyle\frac{2\pi}{w_{\rm ph}}}\,
    \displaystyle\sum\limits_{\alpha=1,2} \mathbf{e}^{(\alpha),*}\;
    \delta(\mathbf{K}_{f} - \mathbf{k})\; 
    \Bigl\langle \Phi_{f}(\mathbf{r}) \Bigl|\,
        Z_{\rm eff}(\mathbf{r})\: e^{-i\,\mathbf{kr}}\: \mathbf{p}\;
      \Bigr|\,
        \Phi_{i}(\mathbf{r})
    \Bigr\rangle,
\end{array}
\label{eq.2.4.7}
\end{equation}
where we introduced the effective charge of the system composed from the $\alpha$ particle and daughter nucleus,
charged form factor of $\alpha$ particle,
and charged form factor of the daughter nucleus as
\begin{equation}
\begin{array}{lcl}
  Z_{\rm eff}(\mathbf{r}) & = &
    e^{i\,\mathbf{kr}}\:
    \Bigl\{
      e^{- i\,c_{A} \mathbf{kr}}\: c_{A}\, Z_{\rm \alpha}(\mathbf{k}) -
      e^{i\,c_{\alpha} \mathbf{kr}}\: c_{\alpha}\, Z_{\rm A}(\mathbf{k})
    \Bigr\},
\end{array}
\label{eq.2.4.4}
\end{equation}
\begin{equation}
\begin{array}{lcl}
  Z_{\rm \alpha} (\mathbf{k}) =
  \Bigl\langle \psi_{\alpha, f} \Bigl|
    \displaystyle\sum\limits_{i=1}^{4} z_{i}\,\displaystyle\frac{m_{\rm p}}{m_{i}}\;
    e^{-i \mathbf{k} \rhobf_{\alpha i} }\,
  \Bigr|\, \psi_{\alpha, i} \Bigr\rangle, &
  Z_{\rm A} (\mathbf{k}) =
  \Bigl\langle \psi_{\rm nucl, f} \Bigl|\,
    \displaystyle\sum\limits_{j=1}^{A} z_{j}\,\displaystyle\frac{m_{\rm p}}{m_{j}}\;
      e^{-i \mathbf{k} \rhobf_{A j} }
  \Bigr|\, \psi_{\rm nucl, i} \Bigr\rangle.
\end{array}
\label{eq.2.4.5}
\end{equation}
%
%
In the first approximation (called as \definition{dipole}) $\exp(i\mathbf{kr}) \to 1$ we have
\begin{equation}
  Z_{\rm eff}^{\rm (dip)} =
  c_{A} Z_{\rm \alpha}(\mathbf{k}) - c_{\alpha} Z_{\rm A}(\mathbf{k}).
\label{eq.2.4.8}
\end{equation}
One can see that in such an approximation the effective charge becomes independent of the relative distance between centers-of-masses of the $\alpha$ particle and daughter nucleus.
In further calculations we shall restrict ourselves by application of the dipole approximation for determination of the effective charge in form of Eq.~(\ref{eq.2.4.8}), while we shall calculate the matrix element without such an approximation. Such a way allows us to take the multipole corrections into account (in contrast to the dipole approach, for example see \cite{Papenbrock.1998.PRLTA}).


\subsection{Electromagnetic form factors of the $\alpha$ particle and daughter nucleus
\label{sec.2.5}}


%

For further calculation of the electromagnetic form factors (\ref{eq.2.4.5}), we need to know the full wave functions before and after emission of the photon (which corresponds to the unperturbed hamiltonian). For such functions, we use the general formula (\ref{eq.2.3.3}), where one-nucleon wave functions are represented in a form of multiplication of the space and  spin-isospin functions.
In this paper we shall assume that the space wave function of one nucleon should determine probability of displacement of this nucleon relative to its most probable spatial position, which is not concentrated in the center-of-mass of the fragment, but on a particular distance.
We develop such a consideration on the basis of the following simple idea:
the most probable positions of nucleons of the $\alpha$ particle (described by the space one-nucleon wave functions) in the ground state should not coincide with the center-of-mass of the $\alpha$ particle.
In such a case, they represent vertexes of the tetrahedron, while the oscillating space wave functions of the first four states give maximal probabilities in the joint center. 
Such a consideration is more naturally extended on the many-nucleon systems, where the most probable positions of nucleons of nucleus in the ground state should be correlated with uniformly distributed density of the nuclear matter (and with its saturation).
We shall take information about the most probable positions of nucleons (i.e., data about radius-vectors $\rhobf_{0,s}$) from other methods.
By such motivations, we reformulate many-nucleon formalism in our previous model applied for the proton-nucleus scattering in \cite{Maydanyuk_Zhang.2015.PRC}.
We shall show below that such a presentation of the space one-nucleon wave function allows more accurate analysis of a dependence of the bremsstrahlung spectra on the size of the emitted $\alpha$ particle.

Thus, we rewrite the vector of position of nucleon with number $s$ relatively to the center-of-mass of the fragment as
\begin{equation}
  \rhobf_{s} = \rhobf_{0,s} + \tilde{\rhobf}_{s},
\label{eq.2.5.1.2}
\end{equation}
where
$\rhobf_{0,s}$ is the radius vector from the center-of-mass of the fragment to the point of the most probable location of nucleon with number $s$,
$\tilde{\rhobf}_{s}$ is the displacement of nucleon relatively to this point of its most probable location.
Thus, we construct the full one-nucleon wave function in the form:
\begin{equation}
  \psi_{\lambda_{s}} (s) =
  \varphi_{\lambda_{s}} (\rhobf_{s} - \rhobf_{s,0})\,
  \bigl|\, \sigma^{(s)} \tau^{(s)} \bigr\rangle,
\label{eq.2.5.1.3}
\end{equation}
where $\lambda_{s}$ denotes number of state of nucleon with number $s$.
Also we shall assume that space function of nucleon is normalized by the condition:
\begin{equation}
  \displaystyle\int |\varphi_{\lambda} (\tilde{\rhobf}_{s})|^{2}\; \mathbf{d} \tilde{\rhobf}_{s} = 1.
\label{eq.2.5.1.4}
\end{equation}
Using a one-nucleon representation for the wave function,
we find for the $\alpha$ particle the following form of the form factor (see Appendix~\ref{sec.app.1}):
\begin{equation}
\begin{array}{lcl}
  Z_{\rm \alpha} (\mathbf{k}) & = &
  \displaystyle\frac{Z_{\rm \alpha}}{4} \;
    \displaystyle\sum\limits_{i=1}^{4} e^{- i\,\mathbf{k} \rhobfsm_{i,0}},
\end{array}
\label{eq.2.5.2.3.1}
\end{equation}
and the form factor for the daughter nucleus obtains the following form:
\begin{equation}
\begin{array}{lcl}
  Z_{\rm d} (\mathbf{k}) & = &
  2\, e^{-\, (a^{2} k_{x}^{2} + b^{2} k_{y}^{2} + c^{2} k_{z}^{2})\,/4}\;
  f_{1}\, (\mathbf{k}, n_{1} \cdots n_{A_{\rm d}})\;
  f_{2}\, (\mathbf{k}, \rho_{1} \cdots \rho_{A_{\rm d}}),
\end{array}
\label{eq.2.5.2.3.2}
\end{equation}
where
\begin{equation}
\begin{array}{lcl}
  f_{1}\, (\mathbf{k}, n_{1} \cdots n_{A_{\rm d}}) \; =
  \hspace{-2mm}
  \displaystyle\sum\limits_{n_{x}, n_{y},n_{z} = 0}^{n_{x} + n_{y} + n_{z} \le N}
  \hspace{-2mm}
    L_{n_{x}} \Bigl[a^{2} k_{x}^{2}/2\Bigr]\:
    L_{n_{y}} \Bigl[b^{2} k_{y}^{2}/2\Bigr]\:
    L_{n_{z}} \Bigl[c^{2} k_{z}^{2}/2\Bigr], \\

  f_{2}\, (\mathbf{k}, \rho_{1} \cdots \rho_{A_{\rm d}}) \; = \;
  \displaystyle\frac{1}{A_{\rm d}}\:
  \displaystyle\sum\limits_{j=1}^{A_{\rm d}}
    e^{- i\,\mathbf{k} \rhobfsm_{j,0}}.
\end{array}
\label{eq.2.5.2.3.3}
\end{equation}
Here, function $f_{1}$ is summation over all states of one-nucleon space wave function, function $f_{2}$ describes space distribution of nucleons inside the nucleus.

\subsection{The effective charge and bremsstrahlung probability
\label{sec.2.5.6}}

Let us calculate the effective charge in the dipole approximation. Substituting formulas (\ref{eq.2.5.2.3.1}) and (\ref{eq.2.5.2.3.2}) for the form factors of the $\alpha$ particle and the daughter nucleus to Eq.~(\ref{eq.2.4.8}),
we obtain:
%
%
%
%
%
\begin{equation}
\begin{array}{lcl}
  \vspace{1mm}
  Z_{\rm eff}^{\rm (dip)} & = &
  2\, e^{-\, (a^{2} k_{x}^{2} + b^{2} k_{y}^{2} + c^{2} k_{z}^{2})\,/4}\:
  \Bigl\{
    c_{A}\:
      f_{2\,\alpha}\, (\mathbf{k}, \rho_{1} \cdots \rho_{4})\; - 

    c_{\alpha}\,
      f_{1,\, {\rm d}}\, (\mathbf{k}, n_{1} \cdots n_{A_{\rm d}})\,
      f_{2,\, {\rm d}}\, (\mathbf{k}, \rho_{1} \cdots \rho_{A_{\rm d}})
  \Bigr\}.
\end{array}
\label{eq.2.5.6.1}
\end{equation}
Now we rewrite the matric element of the photon emission as
\begin{equation}
\begin{array}{lcl}
  \langle \Psi_{f} |\, \hat{H}_{\gamma} |\, \Psi_{i} \rangle & = &
%
  -\,\displaystyle\frac{e}{m_{\rm p}}\, \sqrt{\displaystyle\frac{2\pi}{w_{\rm ph}}} \cdot
    p_{fi}\;
    \delta(\mathbf{K}_{f} - \mathbf{k}),
\end{array}
\label{eq.2.5.6.2}
\end{equation}
where
\begin{equation}
\begin{array}{lcl}
  \vspace{2mm}
  p_{fi} \; = \;
  2\, e^{-\, (a^{2} k_{x}^{2} + b^{2} k_{y}^{2} + c^{2} k_{z}^{2})\,/4}\:
  \displaystyle\sum\limits_{\alpha=1,2} \mathbf{e}^{(\alpha),*} \cdot
    \Bigl\langle \varphi_{f}(\mathbf{r}) \Bigl|\,
      \tilde{Z}_{\rm eff}^{\rm dip}\: e^{-i\,\mathbf{kr}}\: \mathbf{p}\;
    \Bigr|\, \varphi_{i}(\mathbf{r})
    \Bigr\rangle, \\

  \tilde{Z}_{\rm eff}^{\rm (dip)} =
    c_{A}
      f_{2\alpha} (\mathbf{k}, \rho_{1} \cdots \rho_{4}) -
    c_{\alpha}
      f_{1, {\rm d}} (\mathbf{k}, n_{1} \cdots n_{A_{\rm d}})
      f_{2, {\rm d}} (\mathbf{k}, \rho_{1} \cdots \rho_{A_{\rm d}}).
\end{array}
\label{eq.2.5.6.3}
\end{equation}
We define the probability of the emitted photons on the bass of matrix element (\ref{eq.2.5.6.2}) in frameworks of formalism given in \cite{Maydanyuk.2012.PRC} and we do not repeat it in this paper.
In result, we obtain the bremsstrahlung probability as%
\footnote{We obtain the formula (\ref{eq.2.6.1}) in dependence on mass of proton $m_{\rm p}$ while in Ref.~\cite{Maydanyuk.2012.PRC} we had the bremsstrahlung probability (49) in dependence on the reduced mass $\mu$. Such a difference is explained by that in the current paper we develop formalism on the basis of the emission operator of the many-nucleon system (\ref{eq.2.2.3}) while in Ref.~\cite{Maydanyuk.2012.PRC} we started the formalism on the basis of the operator of emission (4) of the proton-nucleus system defined via the reduced mass of proton and nucleus.}

\begin{equation}
\begin{array}{ccl}
  \displaystyle\frac{d^{2}\,P (\theta_{f})}{dw_{\rm ph}\,
  d\cos{\theta_{f}}} & = &
    \displaystyle\frac{e^{2}}{2\pi\,c^{5}}\:
      \displaystyle\frac{w_{\rm ph}\,E_{i}}{m_{\rm p}^{2}\,k_{i}} \;
      \biggl\{ p_{fi}\, \displaystyle\frac{d\, p_{fi}^{*}(\theta_{f})}{d\,\cos{\theta_{f}}}
      + {\rm c. c.} \biggr\},
\end{array}
\label{eq.2.6.1}
\end{equation}
where c.~c. is complex conjugation,
$p_{fi}$ is proportional to the electrical component $p_{\rm el}$ in Eqs.~(10) in \cite{Maydanyuk.2012.PRC}
[with the additional factor of $2\, e^{-\, (a^{2} k_{x}^{2} + b^{2} k_{y}^{2} + c^{2} k_{z}^{2})\,/4}$ and the included effective charge $\tilde{Z}_{\rm eff}^{\rm (dip)}$] and
$d\,p_{fi} (\theta_{f})\, / d\,\cos{\theta_{f}}$ is defined by the same way as $d\,p\, (k_{i}, k_{f}, \theta_{f})\, / d\,\cos{\theta_{f}}$ in Ref.~\cite{Maydanyuk.2012.PRC}.


\section{Calculations and analysis
\label{sec.results}}

\begin{figure}[htbp]
\centerline{\includegraphics[width=84mm]{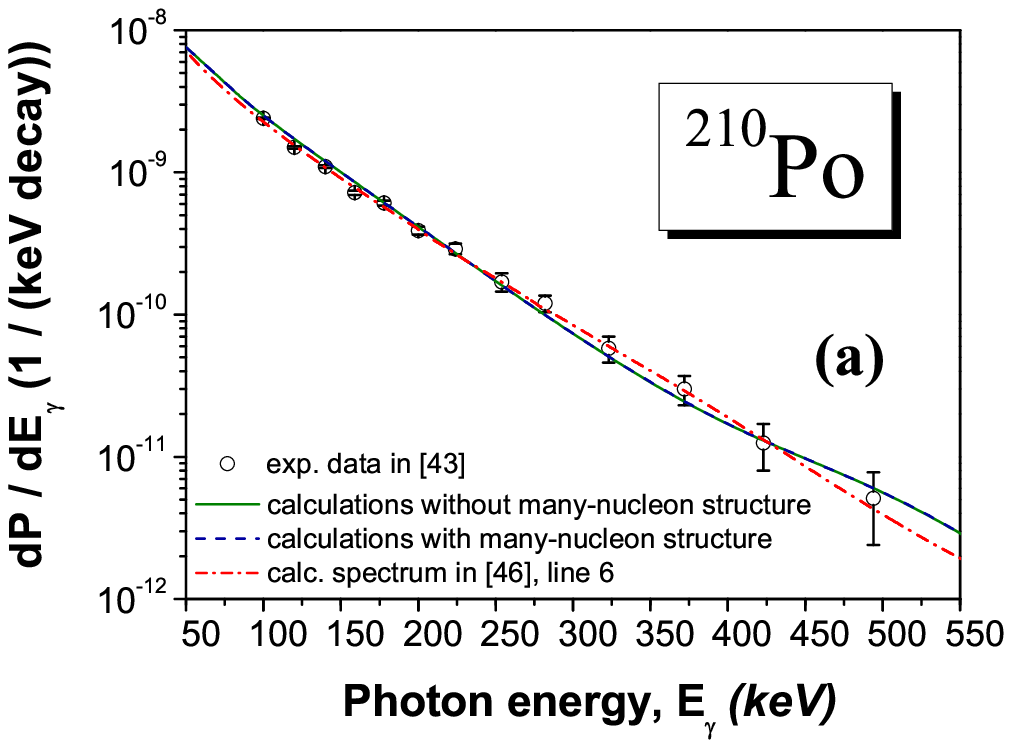}
\hspace{-5mm}\includegraphics[width=84mm]{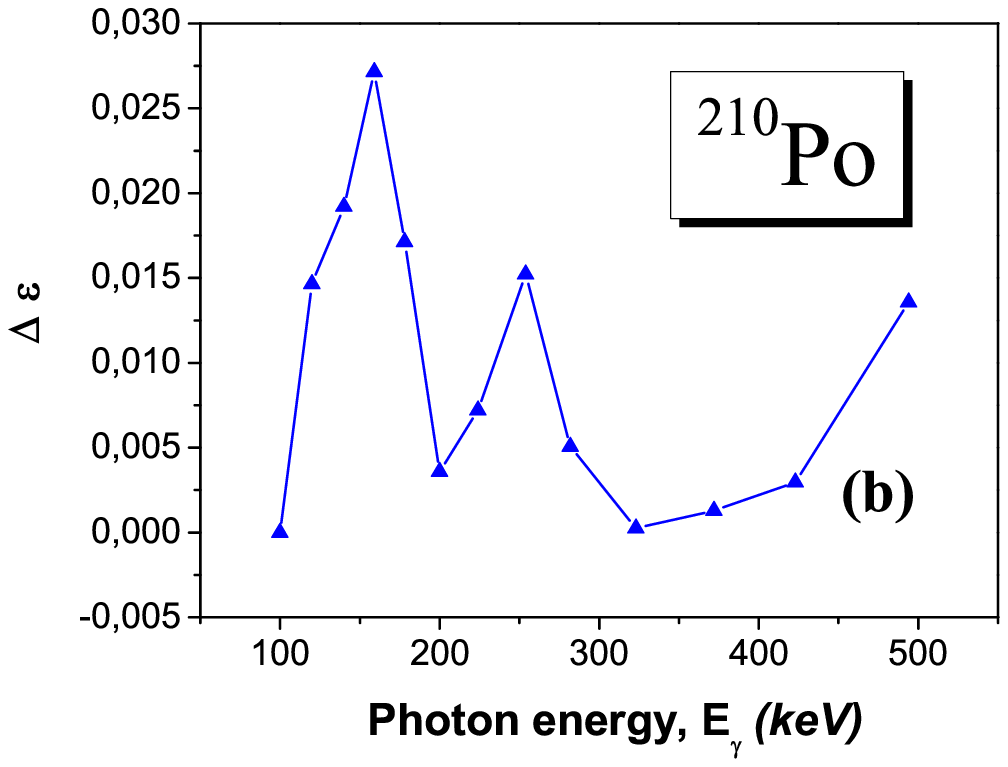}
}
\vspace{-3mm}
\caption{\small 
Figure a:
The bremsstrahlung probabilities of the emitted photons in the $\alpha$ decay of the $^{210}{\rm Po}$ nucleus and experimental data~\cite{Boie.2007.PRL}
[in calculations $\rhobfsm_{i,0}=1.7$~fm for the $\alpha$ particle and $\theta_{f} = 90^{\circ}$ are used,
$\theta_{f}$ is angle between direction of the $\alpha$ particle motion (or its tunneling) after emission of photon and direction of the photon emission].
Here,
solid green line is the calculations without the included nucleon structure,
dashed blue line is the calculations with the included nucleon structure,
dash-dotted red line is calculations for the point-like $\alpha$ particle taken from \cite{Maydanyuk.2006.EPJA} (see line 6 in Fig.~1 in that paper).
The spectrum for calculations with inclusion of the many-nucleon structure is located above the spectrum without this structure,
but such a difference is practically not visible.
Figure b:
The difference of the functions of errors, $\Delta \varepsilon (E_{k})$, defined in Eqs.~(\ref{eq.results.2}) and
obtained after comparative analysis between the new calculations with and without the nucleon structure and experimental data
given in Figure~(a).
\label{fig.1}}
\end{figure}

We apply the method to calculate the spectrum of photons emitted during the $\alpha$ decay.
We started our calculations from the $^{210}{\rm Po}$, $^{214}{\rm Po}$ and $^{226}{\rm Ra}$ nuclei, for which there are experimental data of the bremsstrahlung spectra~\cite{Kasagi.1997.JPHGB,Kasagi.1997.PRLTA,Maydanyuk.2008.EPJA,Giardina.2008.MPLA,D'Arrigo.1994.PHLTA,Boie.2007.PRL}, and our previous developments of the model and results~\cite{Maydanyuk.2003.PTP,Maydanyuk.2006.EPJA,
Maydanyuk.2008.EPJA,Giardina.2008.MPLA,Maydanyuk.2009.NPA,Maydanyuk.2009.TONPPJ,Maydanyuk.2009.JPS} were tested.
Of course, we were initially interested in analysis of the $^{210}{\rm Po}$ nucleus, where the experimental data~\cite{Boie.2007.PRL} were obtained with the best accuracy.
But, the difference between calculations with the included many-nucleon structure and without it is practically not visible [see Fig.~\ref{fig.1}(a)].
Both calculations describe these experimental data enough well.
It turns out that for all these nuclei above, where we have any experimental information, the inclusion of the nucleon structure of the $\alpha$ particle and the daughter nucleus is practically not visible in the bremsstrahlung spectra
(the second digit of the calculated spectrum is varied) for the energy region of the emitted photons below 1 MeV (such a limit is the highest in the experimental data).
However, we find that such an inclusion increases the full bremsstrahlung probability of the emitted photons for each studied nucleus.
This is the first conclusion which we have obtained.

In such a situation, one can add the following analysis.
We define the following functions of errors:
\begin{equation}
\begin{array}{lcl}
  \varepsilon^{\rm (s)} (E_{k}) & = &
    \displaystyle\frac{\Bigl|\ln(\sigma^{\rm (theor,s)} (E_{k})) - \ln(\sigma^{\rm (exp)} (E_{k})) \Bigr|}
    {\Bigl| \ln(\sigma^{\rm (exp)} (E_{1})) \Bigr|},
\end{array}
\label{eq.results.1}
\end{equation}
define the difference between these functions and calculate the summation
\begin{equation}
\begin{array}{lcl}
  \Delta \varepsilon (E_{k}) = \varepsilon^{\rm (no-micro)} (E_{k}) - \varepsilon^{\rm (micro)} (E_{k}), &
  \Delta \bar{\varepsilon} =
    \displaystyle\frac{1}{N} \displaystyle\sum\limits_{k=1}^{N}   \Delta \varepsilon (E_{k}).
\end{array}
\label{eq.results.2}
\end{equation}
%
%
%
Here,
$\sigma^{\rm (theor,s)} (E_{k})$ and $\sigma^{\rm (exp)} (E_{k})$ are
the theoretical and experimental bremsstrahlung probabilities in the $\alpha$ decay at energy $E_{k}$ of the emitted photon,
$s$ is indication of inclustion of the many-nucleon structure into calculations (we denote such calculations by index $micro$) or
calculations without such a many-nucleon structure (we shall use index $no-micro$ for such a case),
and the summation is performed over experimental data.
Such definitions are based on a minimization method (see Ref.~\cite{Maydanyuk_Zhang.2015.NPA}).

%
In order to estimate if inclusion of the many-nucleon structure into calculations provides a better description of the experimental data,
we have to find a difference $\Delta \varepsilon (E_{k})$ between functions
$\varepsilon^{\rm (no-micro)}(E_{k})$ and $\varepsilon^{\rm (micro)}(E_{k})$.
Such calculations for the $\alpha$ decay of the $^{210}{\rm Po}$ nucleus are given in Fig.~\ref{fig.1}(b).
One can see that the function is positive inside the whole photon energy region, that indicates that inclusion of the many-nucleon structure into calculations is more successful in description of the experimental data \cite{Boie.2007.PRL}.
A general estimation can be obtained via the summarizing characteristic in Eqs.~(\ref{eq.results.2}),
and we obtain $\Delta \bar{\varepsilon} = 0.00001015$ (that is positive also).

\begin{figure}[htbp]
\centerline{\includegraphics[width=85mm]{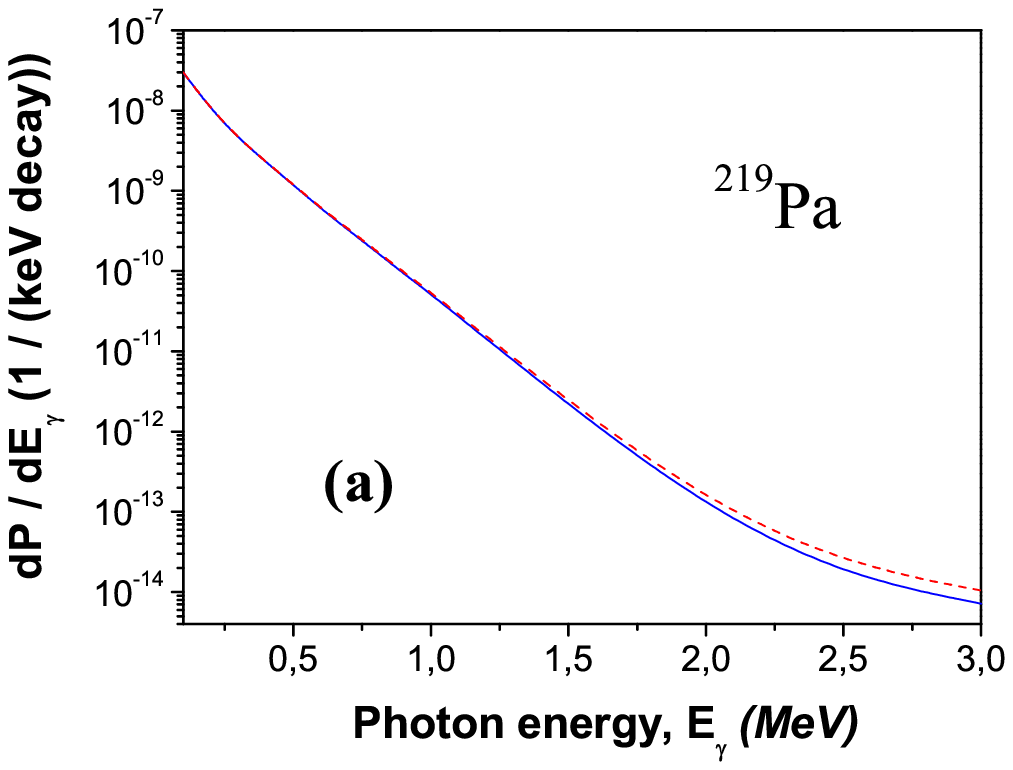}
\hspace{-1mm}\includegraphics[width=85mm]{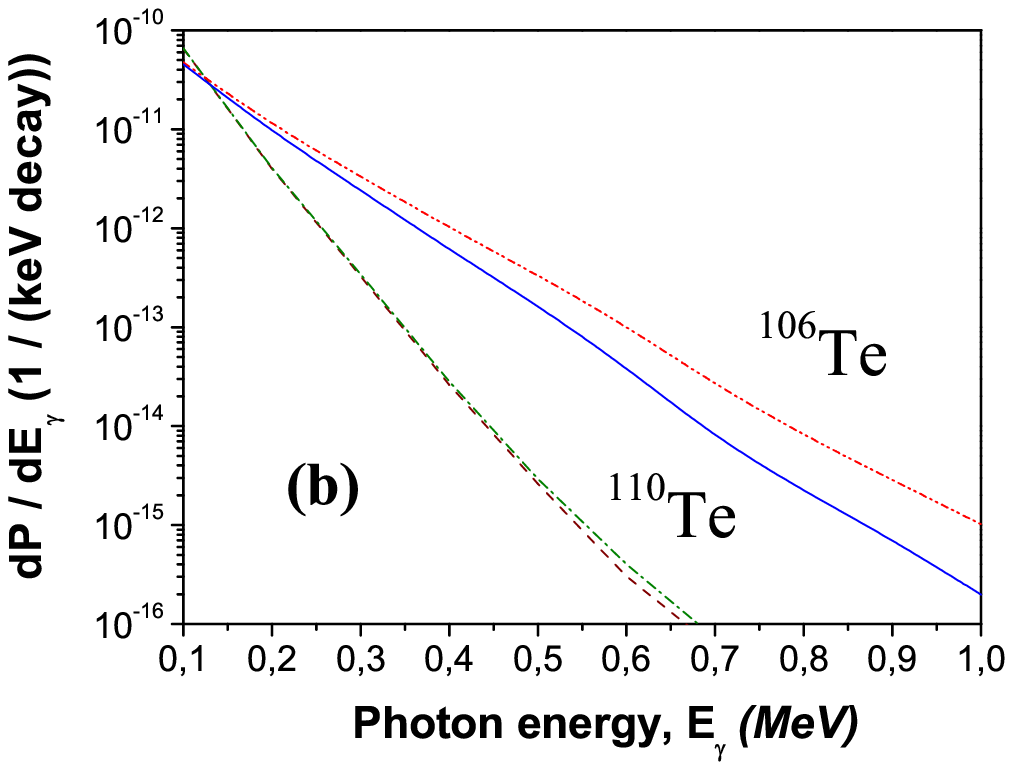}}
\vspace{-3mm}
\caption{\small 
The bremsstrahlung probabilities of the emitted photons at the $\alpha$ decay of the nuclei $^{219}{\rm Pa}$ (a), $^{106}{\rm Te}$ and $^{110}{\rm Te}$ (b)
[in calculations $\rhobfsm_{i,0}=1.7$~fm for the $\alpha$ particle and $\theta_{f} = 90^{\circ}$ are used,
$\theta_{f}$ is angle between direction of the $\alpha$ particle motion (or its tunneling) after emission of photon and direction of the photon emission].
Figure a:
dashed red line for calculations for $^{219}{\rm Pa}$ with included nucleon structure,
solid blue line for calculations for $^{219}{\rm Pa}$ without included nucleon structure,
Figure b:
dash-double dotted red line is for calculations for $^{106}{\rm Te}$ with included nucleon structure,
solid blue line for calculations for $^{106}{\rm Te}$ without nucleon structure,
dash-dotted green line for calculations for $^{110}{\rm Te}$ with included nucleon structure,
dashed brown line for calculations for $^{110}{\rm Te}$ without included nucleon structure.
The curve for each nucleus after inclusion of the nucleon structure is located above in comparison with curve without this structure. The nucleon structure is visible more strongly in the spectra for nuclei with higher $Q_{\alpha}$-values and smaller $Z$.
\label{fig.2}}
\end{figure}

As the next step, we began to search other $\alpha$ decaying nuclei, for which this effect (of influence of the nucleon structure on the bremsstrahlung spectra) could be visible practically.
In selection of appropriate nuclei we have chosen the following basis:
\begin{enumerate}
\item
Possible new measurements of the bremsstrahlung photons will have smaller experimental errors for the $\alpha$ decaying nuclei with higher emitted probabilities (at the same energies of the emitted photons), that corresponds to higher $Q_{\alpha}$-values of these nuclei.

\item
Calculations of the bremsstrahlung spectra are more stable and give more convergent results at higher $Q_{\alpha}$-values of the $\alpha$ decaying nuclei and at lower energies of the emitted photons.
\end{enumerate}

The form factors are included into the effective charge $Z_{\rm eff}$ and, therefore, this effective charge should determine degree of variations of the bremsstrahlung spectrum after inclusion of the nucleon structure into the model and calculations.
However, analyzing the different nuclei, we have found that such spectra variations are enough slow.
Moreover, $Q_{\alpha}$-value influences the probability of the bremsstrahlung photons. This parameter is larger and the emission of photons is more intensive. 
Hence, one can conclude that the nucleon structure should be visible for nuclei with higher $Q_{\alpha}$-values.
This idea allows us to extend the region of our search of proper nuclei, which was our next step. 
In general, $Q_{\alpha}$-value is gradually increased at increasing of the mass of nucleus.
Therefore, we have looked for nuclei in the direction of the heaviest where nucleon structure could be visible in the bremsstrahlung spectra.
However, the next calculations have shown that for heavy and super-heavy nuclei such an influence of the nucleon structure is not visible again [see Fig.~\ref{fig.2}~(a)].

But there are other parameters which play important roles in such a search. In particular, this is the Coulomb barrier determined by the electromagnetic charge of the daughter nucleus. For light nuclei this parameter is essentially smaller and, therefore, the probability of the emitted photons for such nuclei should be larger.
Analyzing distributions of $Q$-values for the $\alpha$ decaying nuclei, we have chosen isotopes of ${\rm Te}$.
Results of our calculations for these nuclei are presented in Fig.~\ref{fig.2}~(b). We have observed that such an visible change of the bremsstrahlung spectra after inclusion of the nucleon structure into the model and calculations is really present for the $^{106}{\rm Te}$ nucleus ($Q_{\alpha}=4.29$~MeV, $T_{1/2}$=70~mks) even for the photons energies below 1~MeV. However, for the $^{110}{\rm Te}$ nucleus with smaller $Q_{\alpha}$-value ($Q_{\alpha}=2.73$~MeV) this visible role of nucleon structure in spectra is already lost. In general, one can see that inclusion of the nucleon structure increases the probability of the emitted photons for all studied nuclei.


\section{Conclusions and perspectives
\label{sec.conclusions}}

In this paper we have studied if the nucleon structure of the $\alpha$ decaying nucleus can be visible in the experimental bremsstrahlung spectra of the emitted photons
which accompany such a decay.
In this regard, we have developed a new formalism which takes into account the distribution of nucleons in the $\alpha$ decaying nuclear system in the model of bremsstrahlung.
Conclusions from analysis on the basis of this model are the following:
\begin{enumerate}
\item
After inclusion of the nucleon structure into the model the calculated bremsstrahlung spectrum is changed very slowly for the majority of the $\alpha$ decaying nuclei [see Fig.~~\ref{fig.2}~(a) for the $\alpha$ decay of $^{219}{\rm Pa}$]. However, we have observed that visible changes really exist for the $^{106}{\rm Te}$ nucleus ($Q_{\alpha}=4.29$~MeV, $T_{1/2}$=70~mks) even for the energy of the emitted photons up to 1~MeV [see Fig.~~\ref{fig.2}~(b)].

\item
Inclusion of the nucleon structure into the model increases the bremsstrahlung probability of the emitted photons.

\item
We find the following tendencies for obtaining the nuclei,
which have the bremsstrahlung spectra more sensitive to the nucleon structure:
(a) direction to nuclei with smaller $Z$,
(b) direction to nuclei with larger $Q_{\alpha}$-values.

\end{enumerate}
One can suppose that the nucleon structure should be more visible in the bremsstrahlung spectra of the emitted photons in cluster decays, fission, scattering of protons and light charged particle off nuclei (energies can be essentially higher, then $Q$-values of the $\alpha$ decaying nuclei). 
This suggests a direction for further research.


\section*{Acknowledgements
\label{sec.acknowledgements}}

S.~P.~M. is grateful to
Dr.~Andrii~I.~Steshenko for his insight and support in understanding the microscopic approaches developments.
S.~P.~M. thanks the Institute of Modern Physics of Chinese Academy of Sciences for its warm hospitality and support.
This work was supported by the Major State Basic Research Development Program in China (No. 2015CB856903), the National Natural Science Foundation of
China (Grant Nos. 11035006 and 11175215), and
the Chinese Academy of Sciences fellowships for researchers from developing countries (No. 2014FFJA0003).


\appendix
\section{The form factors of the $\alpha$ nucleus system
\label{sec.app.1}}

\subsection{Form factor of the $\alpha$ particle
\label{sec.2.5.4}}

We shall calculate the matrix element (\ref{eq.2.4.5}) in form
\begin{equation}
\begin{array}{lcl}
  Z (\mathbf{k}) & = &
%
  \displaystyle\frac{1}{A}
  \displaystyle\sum\limits_{i=1}^{A}
  \displaystyle\sum\limits_{k=1}^{A}\,
    \Bigl\langle \psi_{k}(i)\, \Bigl|\,
      \displaystyle\frac{Z_{k}\,m_{\rm p}}{m_{k}}\;  e^{-i \mathbf{k} \rhobfsm_{i} }\,
    \Bigl|\, \psi_{k}(i)\, \Bigr\rangle,
\end{array}
\label{eq.2.5.1.7}
\end{equation}
%
%
%
%
where we take into account the orthogonality between wave functions
%
$ \langle \psi_{k}(j)\, |\, \psi_{m}(j)\, \rangle = \delta_{mk}$.
%
%
%
Taking into account zero charge of neutron, we sum Eq.~(\ref{eq.2.5.1.7}) over spin-isospin states.
For even-even fragments we obtain:
\begin{equation}
\begin{array}{lcl}
  Z_{\rm A} (\mathbf{k}) & = &
  \displaystyle\frac{2}{A}
  \displaystyle\sum\limits_{i=1}^{A}
  \displaystyle\sum\limits_{k=1}^{B}\,
    \langle \varphi_{k}(\tilde{\rhobf}_{i})\, |\,
      e^{-i \mathbf{k} \rhobfsm_{i}}\,
    |\, \varphi_{k}(\tilde{\rhobf}_{i})\, \rangle,
\end{array}
\label{eq.2.5.2.1}
\end{equation}
where $B$ is the number of states of the space wave function of nucleon.
Taking into account spin-isospin states, we obtain:
%
$B = A / 4 $.
%
In particular, for the $\alpha$ particle we have $B=1$.

We shall choose the space wave function of one nucleon in the gaussian form, according to formalism in Appendix~A in \cite{Maydanyuk_Zhang.2015.PRC}.
Substituting it into Eq.~(\ref{eq.2.5.2.1}), we find the form factor for the $\alpha$ particle:
\begin{equation}
\begin{array}{lcl}
  Z_{\rm \alpha} (\mathbf{k}) & = &
%
%
%
  \displaystyle\frac{1}{2}\;
  \displaystyle\sum\limits_{i=1}^{4}
    I_{x}(n_{x})\, I_{y}(n_{y})\, I_{z}(n_{z}),
\end{array}
\label{eq.2.5.4.1}
\end{equation}
where
\begin{equation}
\begin{array}{lcl}
  I_{x} (n_{x}, x_{i,0}, a) & = &
  N_{\alpha,x}^{2}
  \displaystyle\int
    e^{-\,\frac{(x_{i}-x_{i,0})^{2}}{a^{2}} }\,
    e^{-i\, k_{x} x_{i}}\:
  H_{n_{x}}^{2} \Bigl(\displaystyle\frac{x_{i}-x_{i,0}}{a} \Bigr)\; dx_{i}
\end{array}
\label{eq.2.5.4.2}
\end{equation}
and solutions for $I_{y} (n_{y})$ and $I_{z} (n_{z})$ are obtained after change of indexes $x \to y$ and $x \to z$.
After simplification of this integral we obtain:
\begin{equation}
\begin{array}{lcl}
\vspace{1mm}
  I_{x}
  & = &
  N_{\alpha,x}^{2}\;
    \exp{\Bigl[-\, a^{2} k_{x}^{2}/4 - i\,k_{x} x_{i,0} \Bigr]}\;  
  \displaystyle\int\:
    \exp{\Bigl[-\,\displaystyle\frac{(x_{i}-x_{i,0} + i\,a^{2} k_{x}/2)^{2} }{a^{2}} \Bigr]}\,
    H_{n_{x}}^{2} \Bigl(\displaystyle\frac{x_{i}-x_{i,0}}{a} \Bigr)\; dx_{i}.
\end{array}
\label{eq.2.5.4.3}
\end{equation}

Now let us consider a case when the $\alpha$ particle is in the ground state ($n_{x} = n_{y} = n_{z} = 0$). We have
$H_{n_{x}=0} = 1$,
$H_{n_{y}=0} = 1$,
$H_{n_{z}=0} = 1$.
%
%
In approximation, integral (\ref{eq.2.5.4.3}) over complex variable $\tilde{x} = x_{i}-\rho_{i,x} + i\,a^{2} k_{x}/2$
has solution:
\begin{equation}
\begin{array}{lcl}
  \displaystyle\int\:
      \exp{\Bigl[-\,\displaystyle\frac{(x_{i}-\rho_{i,x} + i\,a^{2} k_{x}/2)^{2} }{a^{2}} \Bigr]}\;
    dx_{i} =
    \displaystyle\int\:
      \exp{\Bigl[-\,\displaystyle\frac{x_{i}^{2}}{a^{2}} \Bigr]}\;
    dx_{i} =
  N_{\alpha,x}^{-2}
\end{array}
\label{eq.2.5.4.5}
\end{equation}
and we obtain:
\begin{equation}
\begin{array}{lcl}
  I_{\alpha,x} (n_{x}=0)
  & = &
  \exp{\Bigl[-\, a^{2} k_{x}^{2}/4 - i\,k_{x} x_{i,0} \Bigr]}.
\end{array}
\label{eq.2.5.4.6}
\end{equation}
Now we calculate form factor (\ref{eq.2.5.4.1}):
%
%
%
%
%
\begin{equation}
\begin{array}{lcl}
  Z_{\rm \alpha} (\mathbf{k}) & = &
  \displaystyle\frac{1}{2}\;
  e^{-\, (a^{2} k_{x}^{2} + b^{2} k_{y}^{2} + c^{2} k_{z}^{2})\,/4}\;
  \displaystyle\sum\limits_{i=1}^{4} e^{- i\,\mathbf{k} \rhobfsm_{i,0}}.
\end{array}
\label{eq.2.5.4.7}
\end{equation}
In limit of point-like $\alpha$ particle (at
$\rhobf_{0,i} = 0$) we obtain:
\begin{equation}
\begin{array}{lcl}
  Z_{\rm \alpha} (\mathbf{k}; \rhobf_{i,0} \to 0) & = &
  2\; e^{-\, (a^{2} k_{x}^{2} + b^{2} k_{y}^{2} + c^{2} k_{z}^{2})\,/4}.
\end{array}
\label{eq.2.5.4.8}
\end{equation}
One can see that the charged form factor depends on the energy of the emitted photon, the direction of its emission, and also on the parameters of the wave function of the nucleon.
In order to make the form factor unambiguous, we impose the following condition:
%
that the form factor of the $\alpha$ particle at point-like limit
should correspond to its electromagnetic charge $Z_{\alpha}$ as
\begin{equation}
  2\, e^{-\, (a^{2} k_{x}^{2} + b^{2} k_{y}^{2} + c^{2} k_{z}^{2})\,/4} \equiv
  Z_{\rm \alpha}
\label{eq.2.5.4.9}
\end{equation}
Applying such a condition, we obtain:
\begin{equation}
\begin{array}{lcl}
  Z_{\rm \alpha} (\mathbf{k}) & = &
  \displaystyle\frac{Z_{\rm \alpha}}{4} \cdot
    \displaystyle\sum\limits_{i=1}^{4} e^{- i\,\mathbf{k} \rhobfsm_{i,0}}.
\end{array}
\label{eq.2.5.4.10}
\end{equation}
On such a basis we construct the following logic.
If the photon is not emitted by the nucleon of the $\alpha$ particle, then $|k|=0$ and we directly obtain fulfilment of property (\ref{eq.2.5.4.9}). However, if the photon is emitted by this nucleon, then the exponent suppresses the form factor of the $\alpha$ particle. This effect is appeared after taking into account of the internal structure of the $\alpha$ particle, composed of four nucleons.

Now, if to remind that $a$, $b$ and $c$ determine space size of localization of the wave function which describes the most probable location of each nucleon inside the $\alpha$ particle, then one concludes : the parameters is larger, the emitted photon suppresses the electromagnetic charge of the $\alpha$ particle stronger.
And the parameters $a$, $b$ and $c$ smaller, the factors in the wave function like $\exp(-x/a)$ are closer to $\delta$-function, and then emission of the photon does not influence charge of the $\alpha$ particle practically.
According to our preliminary estimations, for the $\alpha$ particle for energies of the emitted photon up to 10~MeV, the charge is not changed essentially. 
However, this is not so for high energies (close to 100~MeV and higher) or for heavy ions and nuclei.


\subsection{Form factor of the daughter nucleus
\label{sec.2.5.5}}

In determination of the form factor of the nucleus we have to take into account non-zero states of one-nucleon space wave function. At first, we find the integral $I_{x}(n_{x} \ne 0)$.
Here, one can apply the following formulas of summation of Hermitian polynomials:
%
\begin{equation}
\begin{array}{lcl}
  \displaystyle\frac{(a_{1}^{2} + a_{2}^{2})^{\mu/2}}{\mu!}\,
  H_{\mu} \Bigl( \displaystyle\frac{a_{1}x_{1} + a_{2}x_{2}}{\sqrt{a_{1}^{2} + a_{2}^{2}}}\Bigr) =
  \hspace{-2mm}
  \displaystyle\sum\limits_{m_{1} + m_{2} = \mu}
  \displaystyle\frac{a_{1}^{m_{1}}}{m_{1}!}
  \displaystyle\frac{a_{2}^{m_{2}}}{m_{2}!}
  H_{m_{1}} (x_{1})\, H_{m_{2}} (x_{2}), \\

  \displaystyle\int\limits_{-\infty}^{+\infty}
    e^{-(x-y)^{2}}\, H_{m}(x)\, H_{n}(x)\; dx =
  2^{n} \sqrt{\pi}\, m!\, y^{n-m}\, L_{n}^{n-m} (-2y^{2})
\end{array}
\label{eq.2.5.5.1}
\end{equation}
at $m \le n$ and where $L_{n}^{n-m}$ is generalized Laguerre polynomial.
At $n=m$ we find:
\begin{equation}
\begin{array}{lcl}
  \displaystyle\int\limits_{-\infty}^{+\infty}
    e^{-(x-y)^{2}}\, H_{n}^{2}(x)\; dx =
  2^{n} \sqrt{\pi}\, n!\, L_{n} (-2y^{2}),
\end{array}
\label{eq.2.5.5.2}
\end{equation}
where $L_{n} = L_{n}^{0}$ is Rodrigues polynomial, defined by the Rodrigues formula
\begin{equation}
\begin{array}{lcl}
  L_{n} (x) =
  \displaystyle\sum\limits_{k=0}^{n}
  \displaystyle\frac{(-1)^{k}}{k!}
  \left(
  \begin{array}{c}
    b \\ k
  \end{array}
  \right)
  x^{k}.
\end{array}
\label{eq.2.5.5.3}
\end{equation}
But for computer calculations the following recurrent formula could be more useful:
\begin{equation}
\begin{array}{lcl}
  L_{k+1} (x) =
  \displaystyle\frac{1}{k+1}\,
  \Bigl[(2k+1-x)\, L_{k} (x) - k\, L_{k-1} (x) \Bigr] &
  \mbox{\rm at } k \ge 1,
\end{array}
\label{eq.2.5.5.4}
\end{equation}
where the first two polynomials equals
\begin{equation}
\begin{array}{cc}
  L_{0} (x) = 1, &
  L_{1} (x) = 1 - x.
\end{array}
\label{eq.2.5.5.5}
\end{equation}
Using formula (\ref{eq.2.5.5.2}), we calculate the integral (\ref{eq.2.5.4.3}) for an arbitrary state:
%
%
%
%
%
%
\begin{equation}
\begin{array}{lcl}
  I_{x}
  & = &
  L_{n_{x}} \Bigl[a^{2} k_{x}^{2}/2\Bigr] \cdot
  \exp{\Bigl[-\, a^{2} k_{x}^{2}/4 - i\,k_{x} \rho_{i,x} \Bigr]}
\end{array}
\label{eq.2.5.5.7}
\end{equation}
and calculate the form factor of the daughter nucleus:
\begin{equation}
\begin{array}{lcl}
  Z_{\rm d} (\mathbf{k}) \;\; = \;\;
  \displaystyle\frac{2\, e^{-\, (a^{2} k_{x}^{2} + b^{2} k_{y}^{2} + c^{2} k_{z}^{2})\,/4}}
    {A_{\rm d}}\; 
  \displaystyle\sum\limits_{n_{x}, n_{y},n_{z} = 0}^{n_{x} + n_{y} + n_{z} \le N}\,
    L_{n_{x}} \Bigl[a^{2} k_{x}^{2}/2\Bigr]\:
    L_{n_{y}} \Bigl[b^{2} k_{y}^{2}/2\Bigr]\:
    L_{n_{z}} \Bigl[c^{2} k_{z}^{2}/2\Bigr] \cdot
  \displaystyle\sum\limits_{i=1}^{A_{\rm d}}
    e^{- i\,\mathbf{k} \rho_{i}}.
\end{array}
\label{eq.2.5.5.8}
\end{equation}
This solution can be rewritten as
\begin{equation}
\begin{array}{lcl}
  Z_{\rm d} (\mathbf{k}) & = &
  2\, e^{-\, (a^{2} k_{x}^{2} + b^{2} k_{y}^{2} + c^{2} k_{z}^{2})\,/4}\;
  f_{1}\, (\mathbf{k}, n_{1} \cdots n_{A_{\rm d}})\;
  f_{2}\, (\mathbf{k}, \rho_{1} \cdots \rho_{A_{\rm d}}),
\end{array}
\label{eq.2.5.5.9}
\end{equation}
where
\begin{equation}
\begin{array}{lcl}
  f_{1}\, (\mathbf{k}, n_{1} \cdots n_{A_{\rm d}}) \; =
  \hspace{-2mm}
  \displaystyle\sum\limits_{n_{x}, n_{y},n_{z} = 0}^{n_{x} + n_{y} + n_{z} \le N}
  \hspace{-2mm}
    L_{n_{x}} \Bigl[a^{2} k_{x}^{2}/2\Bigr]\:
    L_{n_{y}} \Bigl[b^{2} k_{y}^{2}/2\Bigr]\:
    L_{n_{z}} \Bigl[c^{2} k_{z}^{2}/2\Bigr], \\

  f_{2}\, (\mathbf{k}, \rho_{1} \cdots \rho_{A_{\rm d}}) \; = \;
  \displaystyle\frac{1}{A_{\rm d}}\:
  \displaystyle\sum\limits_{i=1}^{A_{\rm d}}
    e^{- i\,\mathbf{k} \rho_{i}}.
\end{array}
\label{eq.2.5.5.10}
\end{equation}
Here, function $f_{1}$ is the summation over all states of the one-nucleon space wave function, function $f_{2}$ describes space distribution of nucleons inside the nucleus
(i.e., it characterizes the density of nucleons in the nucleus).





\vspace{5mm}
\begin{thebibliography}{99}
\bibitem{Pluiko.1987.PEPAN}
  V.~A.~Pluyko, V.~A.~Poyarkov
\newblock
  Phys. El. Part. At. Nucl. \textbf{18} (2), 374--418 (1987).

\bibitem{Kamanin.1989.PEPAN}
  V.~V.~Kamanin, A.~Kugler, Yu.~E.~Penionzhkevich, I.~S.~Вatkin, I.~V.~Коруtin,
\newblock
  Phys. El. Part. At. Nucl. \textbf{20} (4), 743--829 (1989).
\bibitem{Amusia_Buimistrov.1987}
  M. Ya. Amusia, V. M. Buimistrov, B. A. Zon, et al.,
\newblock
  \emph{Polaryzed bremsstrahlung emission of particles and atoms}
\newblock
  ({Nauka}, {Moskva}, 1987), 335~p.

\bibitem{Amusia.1990}
  M.~Ya.~Amusia,
\newblock
  \textit{Bremsstrahlung emission}
\newblock
  ({Energoatomizdat}, {Moskva}, 1990), 208~p.

\bibitem{Maydanyuk.2012.book_bremsstrahlung}
  S.~P.~Maydanyuk,
\newblock
  \textit{Nuclear bremsstrahlung: methods of quantum mechanics and electrodynamics in tasks of emission of photons}
\newblock
  ({Palmarium Academic Publishing}, {Saarbr\"{u}cken}, 2012), 148~p.
\bibitem{Kopitin.1997.YF}
  I.~V.~Kopitin, M.~A.~Dolgopolov, T.~A.~Churakova, A.~S.~Kornev,
\newblock
  Phys. At. Nucl. \textbf{60} (5), 776--785 (1997)
  [Rus. ed.:  Yad. Fiz. \textbf{60} (5), 869--879 (1997)].


\bibitem{Nakayama.1989.PRC}
  K.~Nakayama,
\newblock
  Phys. Rev. \textbf{C39} (4), 1475--1487 (1989).

\bibitem{Herrmann.1991.PRC}
  V.~Herrmann, J.~Speth, K.~Nakayama,
\newblock
  Phys. Rev. \textbf{C43} (2), 394--415 (1991).


\bibitem{Liou.1987.PRC}
  M.~K.~Liou, and Z.~M.~Ding,
\newblock
  Phys. Rev. \textbf{C35} (2), 651 (1987).

\bibitem{Liou.1993.PRC}
  M.~K.~Liou, D.~Lin, and B.~F.~Gibson,
\newblock
  Phys. Rev. \textbf{C47}, 973 (1993).

\bibitem{Liou.1995.PLB.v345}
  M.~K.~Liou, R.~Timmermans, and B.~F.~Gibson,
\newblock
  Phys. Lett. \textbf{B345}, 372 (1995).

\bibitem{Liou.1995.PLB.v355}
  M.~K.~Liou, R.~Timmermans, and B.~F.~Gibson,
\newblock
  Phys. Lett. \textbf{B355}, 606(E) (1995).

\bibitem{Liou.1996.PRC}
  M.~K.~Liou, R.~Timmermans, and B.~F.~Gibson,
\newblock
  Phys. Rev. \textbf{C54} (4), 1574--1584 (1996).

\bibitem{Li.1998.PRC.v57}
  Yi~Li, M.~K.~Liou, and W.~M.~Schreiber,
\newblock
  Phys. Rev. \textbf{C57}, 507 (1998).

\bibitem{Li.1998.PRC.v58}
  Yi~Li, M.~K.~Liou, R.~Timmermans, and B.~F.~Gibson,
\newblock
  Phys. Rev. \textbf{C58}, R1880 (1998).

\bibitem{Timmermans.2001.PRC}
  R.~G.~E.~Timmermans, B.~F.~Gibson, Yi~Li, and M.~K.~Liou,
\newblock
  Phys. Rev. \textbf{C65}, 014001 (2001) [15~p.].

\bibitem{Liou.2004.PRC}
  M.~K.~Liou, T.~D.~Penninga, R.~G.~E.~Timmermans, and B.~F.~Gibson,
\newblock
  Phys. Rev. \textbf{C69}, 011001 (2004).

\bibitem{Li.2005.PRC}
  Y.~Li, M.~K.~Liou, and W.~M.~Schreiber,
\newblock
  Phys. Rev. \textbf{C72}, 024005 (2005).

\bibitem{Timmermans.2006.PRC}
  R.~G.~E.~Timmermans, T.~D.~Penninga, B.~F.~Gibson, and M.~K.~Liou,
\newblock
  Phys. Rev. \textbf{C73}, 034006 (2006).

\bibitem{Li.2011.PRC}
  Yi~Li, M.~K.~Liou, W.~M.~Schreiber, and B.~F.~Gibson,
\newblock
  Phys. Rev. \textbf{C84}, 034007 (2011) [10~p.].


\bibitem{Baye.1985.NPA}
  D.~Baye and P.~Descouvemont,
\newblock
  Nucl. Phys. \textbf{A443}, 302--320 (1985).

\bibitem{Baye.1991.NPA}
  D.~Baye, C.~Sauwens, P.~Descouvemont, and S.~Keller,
\newblock
  Nucl. Phys. \textbf{A529}, 467--484 (1991).

\bibitem{Liu.1990.PRC}
  Q.~K.~K.~Liu, Y.~C.~Tang, and H.~Kanada,
\newblock
  Phys. Rev. \textbf{C42} (5), 1895--1898 (1990).


\bibitem{Dohet-Eraly.2011.PRC}
  J.~Dohet-Eraly, D.~Baye,
\newblock
  Phys. Rev. \textbf{C 84}, 014604 (2011).

\bibitem{Dohet-Eraly.2011.JPCS}
  J.~Dohet-Eraly, J.-M.~Sparenberg, and D.~Baye,
\newblock
  J. Phys.: Conf. Ser. \textbf{321}, 012045 (2011).

\bibitem{Dohet-Eraly.2013.JPCS}
  J.~Dohet-Eraly, D.~Baye, and P.~Descouvemont,
\newblock
  J. Phys.: Conf. Ser. \textbf{436}, 012030 (2013).

\bibitem{Dohet-Eraly.2013.PRC}
  J.~Dohet-Eraly, and D.~Baye,
\newblock
  Phys. Rev. \textbf{C 88}, 024602 (2013).

\bibitem{Dohet-Eraly.2013.PhD}
  J.~Dohet-Eraly,
\newblock
  \emph{Microscopic cluster model of elastic scattering and bremsstrahlung of light nuclei},
\newblock
  PhD thesis (Universite Libre De Bruxelles, 2013).


\bibitem{Edington.1966.NP}
  J.~Edington, and B.~Rose,
\newblock
  Nucl. Phys. \textbf{89}, 523 (1966).

\bibitem{Koehler.1967.PRL}
  P.~F.~M.~Koehler, K.~W.~Rothe, and E.~H.~Thorndike,
\newblock
  Phys. Rev. Lett. \textbf{18}, 933 (1967).

\bibitem{Kwato_Njock.1988.PLB}
  M.~Kwato~Njock, M.~Maurel, H.~Nifenecker, J.~Pinston, F.~Schussler, D.~Barneoud, S.~Drissi,
  J.~Kern, and J.~P.~Vorlet,
\newblock
  Phys. Lett. \textbf{B207}, 269 (1988).

\bibitem{Pinston.1989.PLB}
  J.~A.~Pinston, D.~Barneoud, V.~Bellini, S.~Drissi, J.~Guillot, J.~Julien, M.~Kwato~Njock,
  H.~Nifenecker, M.~Maurel, F.~Schussler, and J.~P.~Vorlet,
\newblock
  Phys. Lett. \textbf{B 218}, 128 (1989).

\bibitem{Pinston.1990.PLB}
  J.~A.~Pinston, D.~Barneoud, V.~Bellini, S.~Drissi, J.~Guillot, J.~Julien,
  H.~Nifenecker, and F.~Schussler,
\newblock
  Phys. Lett. \textbf{B 249}, 402 (1990).

\bibitem{Clayton.1992.PRC}
  J.~Clayton, W.~Benenson, M.~Cronqvist, R.~Fox, D.~Krofcheck, R.~Pfaff,
  M.~F.~Mohar,
  C.~Bloch, D.~E.~Fields,
\newblock
  Phys. Rev. \textbf{C45}, 1815 (1992).

\bibitem{Clayton.1991.PhD}
  J.~E.~Clayton,
\newblock
  \emph{High energy gamma ray production in proton induced reactions
  at energies of 104, 145, and 195 MeV},
\newblock
  PhD thesis (Michigan State University, 1991).
\bibitem{Maydanyuk.2012.PRC}
  S.~P.~Maydanyuk,
\newblock
  Phys. Rev. \textbf{C86}, 014618 (2012), arXiv:1203.1498.

\bibitem{Maydanyuk_Zhang.2015.PRC}
  S.~P.~Maydanyuk, P.M. Zhang,
\newblock
  Phys. Rev. \textbf{C91}, 024605 (2015), arXiv:1309.2784.

\bibitem{Nakayama.1986.PRC}
  K.~Nakayama, G.~Bertsch,
\newblock
  Phys. Rev. \textbf{C34} (6), 2190 (1986).

\bibitem{Remington.1987.PRC}
  B.~A.~Remington, M.~Blann, and G.~F.~Bertsch,
\newblock
  Phys. Rev. \textbf{C35} (5), 1720 (1987).


\bibitem{D'Arrigo.1994.PHLTA}
  A.~D'Arrigo, N.~V.~Eremin, G.~Fazio, G.~Giardina, M.~G.~Glotova, T.~V.~Klochko, M.~Sacchi and A.~Taccone,
\newblock
   Phys. Lett. \textbf{B332} (1--2), 25--30 (1994).

\bibitem{Kasagi.1997.JPHGB}
  J.~Kasagi, H.~Yamazaki, N.~Kasajima, T.~Ohtsuki and H.~Yuki,
\newblock
  Journ. Phys. \textbf{G 23}, 1451--1457 (1997).

\bibitem{Kasagi.1997.PRLTA}
  J.~Kasagi, H.~Yamazaki, N. Kasajima, T.~Ohtsuki and H.~Yuki,
\newblock
  Phys. Rev. Lett. \textbf{79} (3), 371--374 (1997).

\bibitem{Boie.2007.PRL}
   H.~Boie, H.~Scheit, U.~D.~Jentschura, F.~K\"{o}ck,
   M.~Lauer, A.~I.~Milstein, I.~S.~Terekhov, and D.~Schwalm,
\newblock
  Phys. Rev. Lett. \textbf{99}, 022505 (2007).

\bibitem{Maydanyuk.2008.EPJA}
  G.~Giardina, G.~Fazio, G.~Mandaglio, M.~Manganaro,
  C.~Sacc\'{a}, N.~V.~Eremin, A.~A.~Paskhalov, D.~A.~Smirnov, S.~P.~Maydanyuk, and V.~S.~Olkhovsky,
\newblock
  Europ. Phys. Journ. \textbf{A36} (1), 31--36 (2008).

\bibitem{Giardina.2008.MPLA}
  G.~Giardina, G.~Fazio, G.~Mandaglio, M.~Manganaro,
  S.~P.~Maydanyuk, V.~S.~Olkhovsky, N.~V.~Eremin, A.~A.~Paskhalov, D.~A.~Smirnov and C.~Sacc\'{a},
\newblock
  Mod. Phys. Lett. \textbf{A23} (31), 2651--2663 (2008),
  arxiv:~0804.2640.
\bibitem{Maydanyuk.2006.EPJA}
  S.~P.~Maydanyuk and V.~S.~Olkhovsky,
\newblock
  Europ. Phys. Journ. \textbf{A28} (3), 283--294 (2006),
\newblock
  nucl-th/0408022.

\bibitem{Papenbrock.1998.PRLTA}
  T.~Papenbrock, G.~F.~Bertsch,
\newblock
  Phys. Rev. Lett. \textbf{80} (19), 4141--4144 (1998),
\newblock
 nucl-th/9801044.

\bibitem{Tkalya.1999.PHRVA}
  E.~V.~Tkalya,
\newblock
  Phys.~Rev. \textbf{C60}, 054612 (1999).

\bibitem{Jentschura.2008.PRC}
  U.~D.~Jentschura, A.~I.~Milstein, I.~S.~Terekhov, H.~Boie, H.~Scheit, and D.~Schwalm,
\newblock
   Phys. Rev. \textbf{C77}, 014611 (2008).
\bibitem{Maydanyuk.2009.NPA}
  S.~P.~Maydanyuk, V.~S.~Olkhovsky, G.~Giardina, G.~Fazio, G.~Mandaglio and M.~Manganaro,
\newblock
  Nucl.~Phys. \textbf{A823}, 3 (2009).

\bibitem{Ploeg.1995.PRC}
  H.~van~der Ploeg, J.~C.~S.~Bacelar, A.~Buda, C.~R.~Laurens, and A.~van~der~Woude,
\newblock
  Phys.~Rev. \textbf{C52}, 1915 (1995).

\bibitem{Kasagi.1989.JPSJ}
  J.~Kasagi, H.~Hama, K.~Yoschida et al.,
\newblock
  Journ. Phys. Soc. Jpn. Suppl. \textbf{58}, 620 (1989).

\bibitem{Luke.1991.PRC}
  S.~J.~Luke, C.~A.~Gossett, R.~Vandenbosch,
\newblock
  Phys. Rev. \textbf{C44} (4), 1548 (1991).

\bibitem{Varlachev.2007.BRASP}
  V.~A.~Varlachev, G.~N.~Dudkin, V.~N.~Padalko,
\newblock
  Bull. Russ. Acad. Sci.: Phys. \textbf{71} (11), 1635--1639 (2007).

\bibitem{Hofman.1993.PRC}
  D.~J.~Hofman, B.~B.~Back, C.~P.~Montoya, S.~Schadmand, R.~Varma, and P.~Paul,
\newblock
  Phys.~Rev. \textbf{C47}, 1103 (1993).

\bibitem{Eremin.2010.IJMPE}
  N.~V.~Eremin, A.~A.~Paskhalov, S.~S.~Markochev, E.~A.~Tsvetkov, G.~Mandaglio, M.~Manganaro,
  G.~Fazio, G.~Giardina and M.~V.~Romaniuk,
\newblock
  Int. J. Mod. Phys. \textbf{E19}, 1183 (2010).

\bibitem{Pandit.2010.PLB}
  Deepak~Pandit, S.~Mukhopadhyay, Srijit~Bhattacharya, Surajit~Pal, A.~De and S.~R.~Banerjee,
\newblock
  Phys. Lett. \textbf{B690} (5), 473--476 (2010).
\bibitem{Maydanyuk.2010.PRC}
  S.~P.~Maydanyuk, V.~S.~Olkhovsky, G.~Mandaglio, M.~Manganaro, G.~Fazio and G.~Giardina,
\newblock
  Phys. Rev. \textbf{C82}, 014602 (2010).
\bibitem{Maydanyuk.2011.JPG}
  S.~P.~Maydanyuk,
\newblock
  Jour. Phys. \textbf{G38} (8), 085106 (2011).

\bibitem{Thomas.1954.PTP}
  R.~G.~Thomas,
\newblock
  Prog. Theor. Phys. \textbf{12}, 253 (1954).

\bibitem{Delion.1992.PRC}
  D.~S.~Delion, A.~Insolia, R.~J.~Liotta,
\newblock
  Phys. Rev. \textbf{C46}, 1346--1354 (1992).

\bibitem{Xu.2006.PRC}
  C.Xu, Z.~Ren,
\newblock
  Phys. Rev. \textbf{C73}, 041301 (2006).

\bibitem{Delion.2013.PRC}
  D.~S.~Delion, R.~J.~Liotta,
\newblock
  Phys. Rev. \textbf{C87}, 041302 (2013).

\bibitem{Silisteanu.2012.ADNDT}
  I.~Silisteanu, A.~I.~Budaca,
\newblock
  At. Dat. Nucl. Dat. Tables \textbf{98}, 1096--1108 (2012).

\bibitem{Ivascu.1990.PEPAN}
  M.~Ivascu, I.~Silisteanu,
\newblock
  Phys. Elem. Part. At. Nucl. \textbf{21}, 1405 (1990).

\bibitem{Lovas.1998.PRep}
  R.~G.~Lovas, et al.,
\newblock
  Phys. Rep. \textbf{294}, 265 (1998).

\bibitem{Hodgson.2003.PRep}
  P.~E.~Hodgson, E.~Betak,
\newblock
  Phys. Rep. \textbf{374}, 89 (2003).
\bibitem{Batkin.1986.SJNCA}
  I.~S.~Batkin, I.~V.~Kopytin and T.~A.~Churakova,
\newblock
  Yad. Fiz. (Sov. Journ. Nucl. Phys.) \textbf{44}, 1454--1458 (1986).

\bibitem{Dyakonov.1996.PRLTA}
  M.~I.~Dyakonov, I.~V.~Gornyi,
\newblock
  Phys. Rev. Lett. \textbf{76} (19), 3542--3545 (1996).

\bibitem{Dyakonov.1999.PHRVA}
  M.~I.~Dyakonov,
\newblock
  Phys. Rev. \textbf{C60}, 037602 (1999).

\bibitem{Bertulani.1999.PHRVA}
  C.~A.~Bertulani, D.~T.~de~Paula and V.~G.~Zelevinsky,
\newblock
  Phys. Rev. \textbf{C60} (3), 031602 (1999) (4 pages).

\bibitem{Takigawa.1999.PHRVA}
  N.~Takigawa, Y.~Nozawa, K.~Hagino, A.~Ono and D.~M.~Brink,
\newblock
  Phys. Rev. \textbf{C59} (2), R593--R597 (1999).

\bibitem{Flambaum.1999.PRLTA}
  V.~V.~Flambaum and V.~G.~Zelevinsky,
\newblock
  Phys. Rev. Lett. \textbf{83}, 3108--3111 (1999).

\bibitem{Tkalya.1999.JETP}
  E.~V.~Tkalya,
\newblock
  Zh. Eksp. Teor. Fiz. \textbf{116}, 390 (1999)
  [Translation: 1999 {\it Sov. Phys. JETP} \textbf{89} 208].

\bibitem{So_Kim.2000.JKPS} 
  W.~So and Y.~Kim,
\newblock
  Journ. Korean Phys. Soc. \textbf{37} (3), 202--208 (2000).

\bibitem{Misicu.2001.JPHGB}
  S.~Misicu, M.~Rizea and W.~Greiner,
\newblock
  Journ. Phys. G \textbf{27}, 993--1003 (2001).

\bibitem{Dijk.2003.FBSSE}
  W.~van~Dijk and Y.~Nogami,
\newblock
  Few-body systems Supplement \textbf{14}, 229--232 (2003).

\bibitem{Maydanyuk.2003.PTP}
  S.~P.~Maydanyuk, V.~S.~Olkhovsky,
\newblock
  Prog. Theor. Phys. \textbf{109} (2), 203--211 (2003).

\bibitem{Ohtsuki.2006.CzJP} 
  T.~Ohtsuki, H.~Yuki, K.~Hirose, T.~Mitsugashira,
\newblock
  Czech. Journ. Phys. \textbf{56},  D391--D398 (2006).

\bibitem{Amusia.2007.JETP}
  M.~Ya.~Amusia, B.~A.~Zon, and I.~Yu.~Kretinin,
\newblock
  JETP \textbf{105} (2), 343--346 (2007).

\bibitem{Maydanyuk.2009.TONPPJ}
  S.~P.~Maydanyuk,
\newblock
  Open Nucl. Part. Phys. J. \textbf{2}, 17--33 (2009) [open access].

\bibitem{Maydanyuk.2009.JPS}
  S.~P.~Maydanyuk,
\newblock
  Jour. Phys. Study. \textbf{13} (3), 3201 (2009).
















\bibitem{Maydanyuk_Zhang.2015.NPA}
  S.~P.~Maydanyuk, P.-M.~Zhang, and S.~V.~Belchikov,
\newblock
  Nucl. Phys. A, 46p. (2015) --- [in press];
  arxiv:1504.00567.
\bibitem{Landau.v3.1989}
  L.~D.~Landau and E.~M.~Lifshitz,
\newblock
  \textit{Kvantovaya Mehanika, kurs Teoreticheskoi Fiziki}
  (Quantum mechanics, course of Theoretical Physics),
\newblock
  Vol.~3 ({Nauka}, {Mockva}, 1989) p.~768 ---
  [in Russian; eng. variant: Oxford, Uk, Pergamon, 1982].

\end{thebibliography}
\end{document}